\documentclass[nohyper]{JHEP3}

\usepackage{amsmath}
\usepackage{amssymb}
\usepackage{cite}

\usepackage{epsfig}

\psfull

\newcommand{\ie}{\emph{i.e.}\ }
\newcommand{\eg}{\emph{e.g.}\ }

\def\order#1{{\cal{O}}\left(#1\right)}

\def\cS{{\cal{S}}}    
\def\cP{{\cal{P}}}    
\def\cA{{\cal{A}}}    

\def\CF{C_F}

\def\CA{C_A}
\def\NC{N_C}
\def\nf{n_{\!f}}

\def\as{\alpha_{{\textsc{s}}}}

\def\ee{e^+e^-}

\title{Accounting for coherence in interjet $\boldsymbol{E_t}$ flow: a case
  study}

\author{Mrinal Dasgupta \\
  DESY, Theory Group, Notkestrasse 85, Hamburg, Germany.}
\author{Gavin P. Salam \\
LPTHE, Universit\'es P. \& M. Curie (Paris VI) et Denis Diderot
  (Paris VII), Paris, France.}

\abstract{Recently, interest has developed in the distribution of
  interjet energy flows, with for example the
  leading-log calculation of the highly non-trivial colour structure
  of primary emissions in 4-jet systems.  Here we point out however
  that at leading-log level it is insufficient to consider only
  multiple primary emission from the underlying hard antenna ---
  additionally, one must take into account the coherent structure of
  emission from arbitrarily complicated ensembles of large-angle soft
  gluons.  Similar considerations apply to certain definitions of
  rapidity gaps based on energy flow. We examine this new class of
  terms in the simpler context of 2-jet events, and discover features
  that point at novel aspects of the QCD dynamics.
  }


\preprint{DESY--02--021 \\
  LPTHE--02--013\\
  hep-ph/0203009 \\
  February 2002
  }

\begin{document}

\section{Introduction}

The study of interjet transverse energy ($E_t$) flow away from jets
was suggested by Marchesini \& Webber \cite{MW} as a method of
separating QCD Bremsstrahlung contributions from those of the
underlying event in hadron-hadron collisions. Recently, in the
$2+2$-jet case their considerations on mean energy flows into
specified detector regions have been extended by Berger, K\'ucs and
Sterman \cite{BKS} to energy-flow \emph{distributions}.

This work apart from its potential phenomenological value at hadron collider
experiments also represents theoretical advances in understanding and
computing the complex colour topology dictating the flow of soft
gluons in $4$ jet events, and follows in part from earlier work which
discussed perturbative calculations for rapidity gaps in terms of
interjet energy flows \cite{OdeSter,Oderda}.

One of the main aims of Refs.~\cite{BKS,OdeSter,Oderda} is to resum
logarithms $L$ in transverse energy at single-logarithmic accuracy,
\ie all terms $(\as L)^n$. The authors consider the structure of
multiple independent soft emissions, and their virtual corrections,
from an antenna consisting of the four hard 
partons (the incoming and outgoing jets) and show how it
exponentiates. This is a rather delicate procedure because of the
interferences that arise in the colour algebra when squaring the
amplitudes involving  4-hard partons and 
an arbitrary number of soft gluons. The above mentioned 
studies therefore represent a considerable 
advance in the field in terms of understanding the `colour content' of a 
multi-parton hard scattering (see e.g the discussion in Ref.~\cite{OdeSter}). 

\FIGURE{
    \epsfig{file=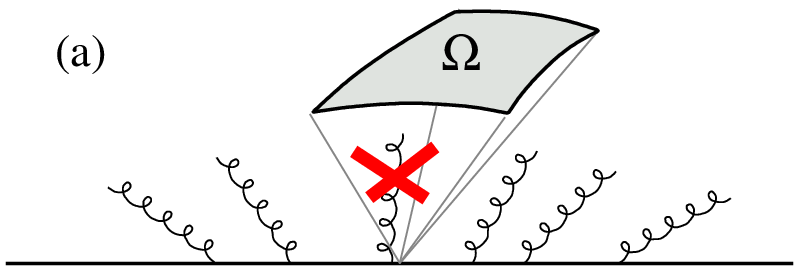,width=0.325\textwidth}
    \epsfig{file=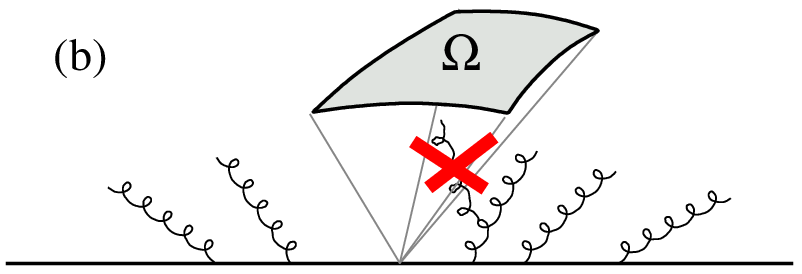,width=0.325\textwidth}
    \epsfig{file=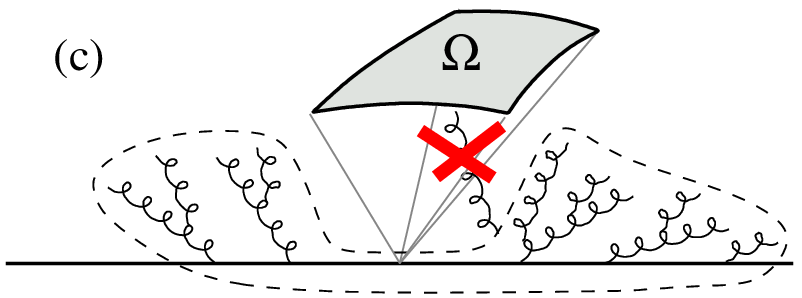,width=0.325\textwidth}
    \caption{(a) veto on primary emissions going into $\Omega$; (b)
      veto on energy-ordered secondary emission going into $\Omega$;
      (c) veto on emissions going into $\Omega$ that are radiated
      coherently from the ensemble made up of all (much) harder
      emissions not going into $\Omega$.}
    \label{fig:SpotTheDifference}
}

In certain contexts, for example for the invariant mass distribution
of a dijet pair \cite{KOSpair}, this exponentiation of what we shall label 
as \emph{primary}
emissions is sufficient. However when considering energy flows in
restricted angular regions precisely as in Refs.~\cite{BKS}, 
or equivalently definitions of diffraction
based on energy cuts \cite{OdeSter,Oderda} (or multiplicity cuts),
there is another class of 
single-logarithmic (SL) terms which must be accounted for.  To be more
specific it was shown in Ref.~\cite{dassalNG} that for any observable
that is sensitive to radiation in only a part of phase space --- a
\emph{non-global} observable --- it is necessary to account for secondary
and higher emissions (as defined below) 
to obtain an answer correct to SL accuracy.\footnote{Another context in
  which non-global terms would be relevant is if one were to attempt
  to carry out a resummation to single-logarithmic accuracy for
  certain kinds of isolation criteria for isolated photons. However we
  are not aware of any such resummation being currently in existence.}

To illustrate this point better we turn to figure
\ref{fig:SpotTheDifference}, which for 
simplicity refers to a $2$-jet event, though the logic remains the same for
the case with several hard jets. It shows a patch $\Omega$ in
rapidity and azimuth into which the total (transverse) energy flow is
restricted to be less than some small amount $Q_\Omega$. There are
various sources of SL terms. One source comes from vetoing `primary'
emissions which fly directly into $\Omega$, 
figure~\ref{fig:SpotTheDifference}a, where primary means that their
radiation pattern corresponds to that for the antenna associated with
the two hard jets.
The lowest-order contribution of this kind comes from an
incomplete cancellation between real and virtual terms and gives $-
\frac{2\as\CF}{\pi} \cA_\Omega \ln \frac{Q}{Q_{\Omega}}$, where $Q$ is
the hard scale of the problem and $\cA_\Omega$ relates to the area of
$\Omega$.

A second source of SL terms comes from diagrams such as
figure~\ref{fig:SpotTheDifference}b. Here we have a
large-angle\footnote{Strictly speaking, large-angle means of the same
  order as the 
  angles involved in the definition $\Omega$ --- if $\Omega$ has a
  boundary at a small angle to a hard-parton axis, then large-angle
  can actually mean `of the same order as that small angle.'} 
primary emission which flies outside $\Omega$, with energy $Q_1$, such
that $Q \gg Q_1 \gg Q_\Omega$ --- this gives us one power of $\as \ln
\frac{Q}{Q_\Omega}$. Forbidding it from radiating a secondary emission
into $\Omega$ gives powers of $\as \ln \frac{Q_1}{Q_\Omega}$, which
after integration over $Q_1$ translate into a set of SL terms, $(\as
\ln \frac{Q}{Q_\Omega})^n$. This
kind of term has been neglected in \cite{BKS,OdeSter,Oderda}, as well as in
several other contexts \cite{ADS,BG,BDMZ}. To correctly account for it
at all orders it is necessary to consider soft emission into $\Omega$ which
is \emph{coherently} radiated from arbitrary ensembles of soft (but
harder), large-angle  
energy ordered gluons outside of $\Omega$,
figure~\ref{fig:SpotTheDifference}c, rather than just the hard 
initiating jets in the picture.  
For simplicity we call
this kind of emission a secondary emission, though this is only a
figure of speech,
since it is
coherently radiated from external ensembles which may consist of
primary, secondary, tertiary, etc.\ gluons. 
In general we refer to the class of terms generated by such
contributions as non-global logarithms. 

In earlier work \cite{dassalNG} we calculated such terms for
observables sensitive to radiation in a single hemisphere of a two-jet
event. This was done exactly to second order in $\as \ln
\frac{Q}{Q_\Omega}$, and numerically in the large-$\NC$ limit at all
orders.  Here we extend this work to the case of interjet energy flow
in two-jet events. We view this in part as an intermediate step to a
calculation in the $3$ and $4$-jet cases, however it also has value in
its own right. Firstly it will turn out that an analysis of the
dependence of the effect on the geometry of the patch $\Omega$ casts
considerable light on the dynamical mechanisms involved in non-global
effects. Secondly it allows us to make a general order of magnitude estimate
of the importance of non-global terms relative to those from the
resummation of primary emissions. Finally the measurement of
energy-flow distributions in 2-jet events in $e^+e^-$ collisions or
DIS could well be of intrinsic interest since it would be
complementary to measurements in hadron-hadron collisions, and in
particular, free of the problems associated with the underlying event.

\section{Primary emission form factor}

In this paper we shall be considering as our observable the amount of
transverse energy $E_t$ flowing into a patch $\Omega$ in rapidity and
azimuth:
\begin{equation}
  E_t = \sum_{i\in \Omega} E_{t,i} \,.
\end{equation}
We are interested in the probability $\Sigma_\Omega$ for $E_t$ to be less
than some value $Q_\Omega$ which is much smaller than the hard scale
$Q$ of the process in question:
\begin{equation}
  \Sigma_\Omega(Q_\Omega, Q) = \frac1\sigma \int_0^{Q_\Omega} dE_t\,
  \frac{d\sigma}{dE_t}\,,
\end{equation}
where $\sigma$ is the Born-order cross section for the process --- in
our case the production of two jets in $\ee$ or of $1+1$ jets in DIS. 

In order, later on, to quantify the effect of non-global logs it is
useful first to calculate the contribution to $\Sigma_\Omega$ from
primary emissions alone. This is the much simpler $2$-jet analogue of
what has been calculated in \cite{BKS} for 4-jet systems.

At first order in $\as$, the logarithmically enhanced contribution to
$\Sigma_\Omega$ comes from the incomplete cancellation of real and
virtual contributions for a soft primary emission:
\begin{equation}
  \label{eq:Sigma1}
  \Sigma^{(1)}_\Omega(Q_\Omega, Q) = - 4 C_F
  \frac{\alpha_s}{2\pi} \int_{Q_\Omega}^{Q/2} \frac{d k_t}{k_t}
  \int_{\Omega} d\eta \frac{d\phi}{2\pi} = 
  -\frac{4\CF \as}{2\pi} \,\cA_\Omega\, \ln \frac{Q}{2Q_\Omega}\,,
\end{equation}
where we have introduced the notation $\cA_\Omega$ for \emph{area} of
the region $\Omega$,
\begin{equation}
  \label{eq:AOmega}
  \cA_\Omega = \int_\Omega d\eta\,\frac{d\phi}{2\pi}\,.
\end{equation}
The upper limit in the $k_t$ integral is arbitrary to single-log
accuracy, as long as it is of order $Q$.  

When the logarithm of
$Q/Q_\Omega$ becomes large enough to compensate the smallness of
$\as$, it is necessary to include terms $(\as \ln
\frac{Q}{Q_\Omega})^n$ to all orders.
If one assumes (incorrectly, as we shall see) that multiple wide-angle
soft gluons from a two-jet system are simply radiated independently according
to a two-particle antenna pattern,
then eq.~\eqref{eq:Sigma1} can be extended to all orders by accounting
for the running of the coupling\footnote{Strictly speaking the running
  of the coupling is connected with the collinear branching
  of the primary gluons. This however is a separate issue from that of
  large-angle soft gluon emission with which we deal later on in this
  article.}
and then exponentiating the answer:
\begin{equation}
  \label{eq:SigmaOmegaP}
  \Sigma_{\Omega,\cP}(Q_\Omega,Q) \equiv 
  \Sigma_{\Omega,\cP}\left(t(Q_\Omega,Q)\right) = 
  \exp \left [ -4 C_F \cA_\Omega t 
 \right ]\,.
\end{equation}
The subscript $\cP$ on $\Sigma_{\Omega,\cP}$ serves as a reminder
that we have only taken into account primary emissions and 
$t$ is defined to be the following integral of $\as$,
\begin{equation}
  \label{eq:tdef}
  t(Q_\Omega,Q) = \frac{1}{2 \pi} 
  \int_{Q_\Omega}^{Q/2} \frac{dk_t}{k_t} \alpha_s(k)
  = \frac{1}{4\pi\beta_0} \ln \frac{\as(Q/2)}{\as(Q_\Omega)}\,,
\end{equation}
where the second equality holds
at the one-loop level and $\beta_0 = (11\CA - 2\nf)/(12\pi)$.

\section{Leading order calculation of non-global effects}

As well as dealing with primary emissions, it is necessary to account
also for contributions from (secondary) emissions coherently radiated
into $\Omega$ from large-angle soft-gluon ensembles outside of
$\Omega$.  We will denote the contribution from such non-global terms
by the function $\cS(t)$, such that to SL accuracy
\begin{equation}
  \Sigma_{\Omega}(t(Q_\Omega,Q)) \equiv \cS(t)\,
  \Sigma_{\Omega,\cP}(t)\,. 
\end{equation}
To start with, we calculate the leading order contribution to $\cS$,
\ie $\cS_2$, where we define the following series expansion for $\cS$:
\FIGURE{\epsfig{file=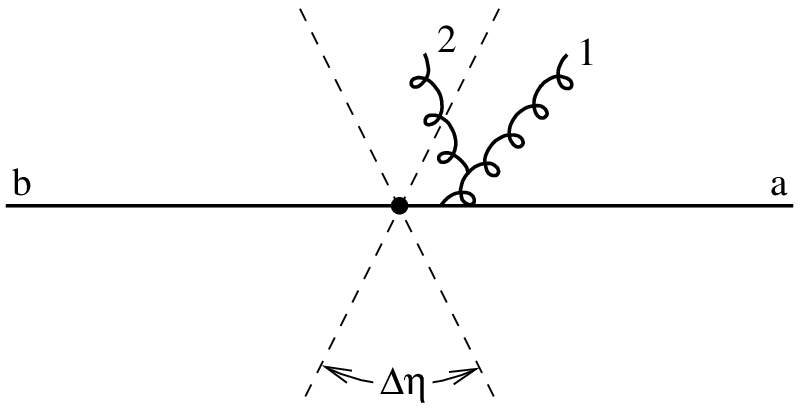,width=0.4\textwidth}
  \caption{The kind of diagram to be considered for the
    calculation of $\cS_2$ in the case of a rapidity slice of width
    $\Delta \eta$.}
  \label{fig:S2diag}
  }
\begin{equation}
  \cS(t) = \sum_{n=2} \cS_n t^n\,.
\end{equation}
Since this kind of contribution only starts with secondary emissions,
there is no $\cS_1$ term. In the calculation of $\cS_2$, we shall be
entitled to equate $t$ with $\frac{\as}{2\pi} \ln
\frac{Q}{2Q_\Omega}$.

The exact value of $\cS_2$ depends on the geometry of the patch
$\Omega$. Here we calculate it analytically for the case where
$\Omega$ is a slice in rapidity of width $\Delta \eta$. The kind of
diagram to be considered is shown in figure~\ref{fig:S2diag}, where $a$
and $b$ are quarks (they may be outgoing or incoming
depending on whether for example we are dealing with $\ee$ or DIS in
the Breit frame) and $1$ and $2$ are gluons. We introduce the
following four-momenta
\begin{subequations}
\begin{align}
  k_a &= \frac{Q}{2}(1,0,0,1)\,, \\
  k_b &=  \frac{Q}{2}(1,0,0,-1)\,, \\
  k_1 &= x_1\frac{Q}{2}(1,0,\sin\theta_1,\cos\theta_1)\,,\\
  k_2 &= x_2\frac{Q}{2} (1,\sin\theta_2 \sin\phi,\sin\theta_2
  \cos\phi,\cos\theta_2)\,,
\end{align}
\end{subequations}
where we have defined energy fractions $x_{1,2} \ll 1$ for the two gluons.
To our accuracy, we can neglect the recoil of the hard particles
against the soft gluons.

We write the squared matrix element for energy-ordered two-gluon emission as
(see for example \cite{DMO})
\begin{subequations}
\begin{align}
  W &= 4\CF \frac{(ab)}{(a1)(1b)} \left( \frac{\CA}{2}
    \frac{(a1)}{(a2)(21)} + \frac{\CA}{2} \frac{(b1)}{(b2)(21)} +
    \left(\CF - \frac{\CA}{2}\right)\frac{(ab)}{(a2)(2b)} \right),
  \\
  &= C_F^2 W_1 + C_F C_A W_2\,,
\end{align} 
\end{subequations}
where $(ij) = k_i \cdot k_j$. The result is valid for $1 \gg x_1 \gg
x_2$ as well as for the opposite ordering of the gluons, and in
addition is completely symmetric under interchange of $k_1$ and $k_2$.
(We have however chosen to write it in an asymmetric form so as to
emphasise the dipole structure of the emissions, namely radiation of
gluon $k_1$ from the $ab$ dipole, followed by the radiation of gluon
$k_2$ from the $a1$, $1b$ and $ab$ dipoles).

The $\CF^2$ piece of the matrix element, $W_1$ corresponds to
independent gluon emission and is included in the primary emission
form factor. To
study specifically the modification relative to the primary emission
case, at this order one must consider the $C_F C_A$
part of the emission probability, $W_2$.

For a general region $\Omega$, $\cS_2$ is defined through the
following equation:
\begin{multline}
  \label{eq:S2DefGen}
  \cS_2 \,\ln^2\! \frac{Q}{2Q_\Omega} + \order{\ln\! \frac{Q}{Q_\Omega}} =
  -C_F C_A
  \int_{k_1 \notin \Omega}\!\!
  d\cos \theta_1 \frac{d\phi_1}{2\pi}
  \int_{k_2 \in \Omega}\!\!
  d\cos \theta_2 \frac{d\phi_2}{2\pi}
  \\ 
\frac{Q^4}{16}\int_{0}^1 x_2 dx_2
\int_{x_2}^{1} x_1 dx_1\, \Theta
 \left (x_2 - \frac{2 Q_\Omega}{Q} \right) W_2\,,
\end{multline}
which takes into both virtual and real contributions.  In the case of
a slice of width $\Delta \eta$, the angular integrals can be rewritten 
explicitly
\begin{multline}
  \int_{k_1 \notin \Omega}\!\!
  d\cos \theta_1 \frac{d\phi_1}{2\pi}
  \int_{k_2 \in \Omega}\!\!
  d\cos \theta_2 \frac{d\phi_2}{2\pi}\\
  \to
  \left( \int_{-1}^{-c} d\!\cos{\theta_1} + \int_{c}^1
    d\!\cos{\theta_1} \right)
  \int_{-c}^{c}  d\!\cos{\theta_2} 
\int_0^{2\pi}\frac{d\phi_1}{2\pi} 
\int_0^{2\pi}\frac{d\phi_2}{2\pi} 
\end{multline}
where we have centred the slice at $\eta=0$ (the results are
independent of its position) and defined $\pm c$ to be the cosines of
the polar angles delimiting the slice,
\begin{equation}
  \Delta \eta = \ln \frac{1+c}{1-c}\,.
\end{equation}
The integrals over the energy fractions in \eqref{eq:S2DefGen} are
straightforward. Keeping only the leading-logarithmic piece,
exploiting the symmetry in $\theta_1 \leftrightarrow \pi-\theta_1$ and
performing the azimuthal average we have
\begin{equation}
  \label{eq:S2AzimuthOnly}
\cS_2 = -
4 C_F C_A 
\int_{-1}^{-c} d\!\cos\theta_1
\int_{-c}^{c} d\!\cos\theta_2  \,
F_2(\cos\theta_1,\cos\theta_2)\,,
\end{equation}  
where the angular function $F_2$ is 
\begin{equation}
  \label{eq:F2}
  F_2 =  \frac{2}{(\cos\theta_2 -
    \cos\theta_1)(1-\cos\theta_1)(1+\cos\theta_2)}\,.
\end{equation}
Integrating over the polar angles we obtain
\begin{multline}
  \label{eq:S2c}
 \cS_2 = -4\CF \CA \left[ \frac{\pi^2}{12} + 
    \ln^2 \frac{1+c}{1-c} - \ln \frac{1+c}{1-c} \ln
    \left(\left(\frac{1+c}{1-c}\right)^2 -1\right)\right. \\ \left.
    -\frac12 \mathrm{Li}_2\left(\left(\frac{1-c}{1+c}\right)^2\right)
    -\frac12 \mathrm{Li}_2\left(1-\left(\frac{1+c}{1-c}\right)^2\right)
  \right],
\end{multline}
which can be expressed in terms of $\Delta \eta$ as follows:
\begin{equation}
  \label{eq:S2eta}
 \cS_2 = -4\CF \CA \left[ \frac{\pi^2}{12} + 
    \left (\Delta \eta \right)^2 - \Delta \eta \ln
    \left(e^{2\Delta \eta} -1\right) 
    -\frac12 \mathrm{Li}_2\left(e^{-2\Delta \eta}\right)
    -\frac12 \mathrm{Li}_2\left(1-e^{2\Delta \eta}\right)
  \right],
\end{equation}
where the dilogarithm function is defined as
\begin{equation}
  \mathrm{Li}_2(z) = \int_z^0 \frac{\ln(1-t)}{t}dt\,.
\end{equation}
\FIGURE{
    \epsfig{file=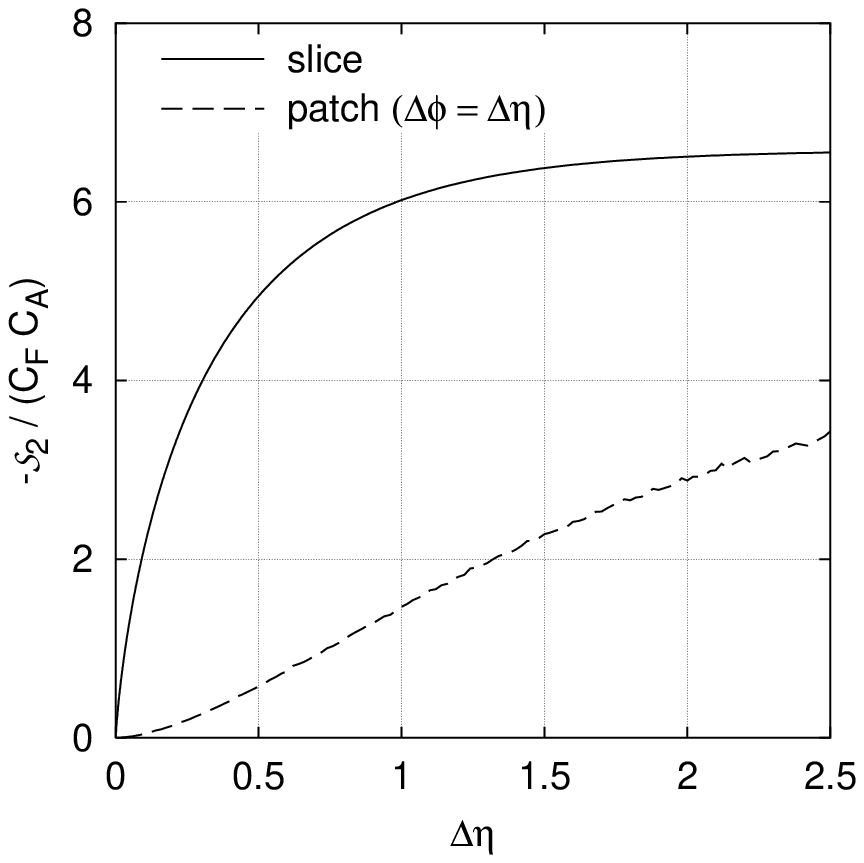,width=0.45\textwidth}
    \caption{$\cS_2$ as a function of $\delta \eta$ for two different
      definitions of $\Omega$: a slice in rapidity (using
      eq.~\eqref{eq:S2eta}), and a square patch in rapidity and azimuth
      with $\delta \phi = \delta\eta$ ($\cS_2$ determined numerically).}
    \label{fig:S2}
}
The functional dependence of $\cS_2$ on $\Delta \eta$ is shown in
figure~\ref{fig:S2}.  
For small $\Delta \eta$, $\cS_2$ goes to zero essentially linearly in
$\eta$,
\begin{multline}
 \cS_2 = -4\CF\CA \left[2(1 - \ln 2\Delta \eta)\Delta \eta 
   + \Delta \eta ^2 
  \right. \\  \left.
    +\order{\Delta \eta^3}\right],
\end{multline}
with a logarithmic enhancement due to the integrable divergence in
eq.~\eqref{eq:S2AzimuthOnly} when $\theta_1 \simeq \theta_2 \simeq
-c$ --- thus $\cS_2$ is roughly proportional to the area
of the slice. On the other hand as $\Delta \eta$ increases, $\cS_2$
rapidly saturates at its asymptotic value,
\begin{equation}
  \lim_{\Delta \eta\to \infty}\cS_2 = -\CF\CA\frac{2\pi^2}{3}\,.
\end{equation}
There is a simple physical reason for this behaviour: $\cS_2$ is
associated with the difference between full coherent emission for a
pair of gluons, and simple independent emission. The dominant
contribution to $\cS_2$ comes therefore from the region where the two
gluons are close together (which by definition means the edges of
$\Omega$ since one gluon is in, while the other is out). On the other
hand, when the two gluons are widely separated in rapidity then
independent emission becomes a good approximation and there is no
contribution to $\cS_2$ --- hence for large $\Delta \eta$, $\cS_2$
receives no contribution from the centre of the slice, only from its
edges, and the value of $\cS_2$ saturates.

As well as showing $\cS_2$ for a slice, figure~\ref{fig:S2} also shows
it (determined numerically) for a square patch in rapidity and
azimuth. Over the range of 
$\Delta\eta = \Delta \phi$ shown, the behaviour is quite different,
with an approximate linearity in $\Delta \eta$ --- this too can be
understood from the above arguments: since it is the edges of $\Omega$
which contribute dominantly to $\cS_2$, the value of $\cS_2$ will be
roughly proportional to the perimeter of the patch, and hence
linear in $\Delta \eta$.  This holds however only for moderate patch
sizes: for very small patches, $\cS_2$ is roughly proportional to the
patch area (with a logarithmic enhancement of similar origin to that for the
slice), while for 
large patches the periodicity in $\phi$ means that the patch tends
to a slice.

\section{All-orders treatment}

While for the $2$-gluon case the analytical calculations are
relatively straightforward, in the many-gluon case the situation is
vastly more complex: not only does emission into $\Omega$ come from a
`Mexican-cactus'-like external multi-gluon ensemble, but the evolution
with $t$ of the ensemble outside $\Omega$ depends itself on the
structure developed at smaller $t$. Difficulties come from both the
geometry and the colour structure of the ensemble. As a result, we
have not succeeded in obtaining even approximate analytical results,
and we have to resort to the large-$\NC$ approximation and numerical
methods in order to extend our calculations to the all-orders case.

\subsection{Possible underlying dynamics}
\label{sec:dynamics}

Before going on to the details of the numerical results it is however
instructive to consider some very rough arguments concerning the
underlying dynamics. One way in which $\Omega$ can stay empty is
simply to prevent all emitters outside of $\Omega$ from emitting into
$\Omega$. For each emitter `close' to the edge of $\Omega$ the price
to pay is roughly $e^{-t}$ (we ignore the coefficient of $t$); since,
to a first approximation, the typical number of relevant emitters will
be of order $t$, we end up with a suppression that goes as $e^{-t^2}$.

Of course when considering very rare configurations, it is risky to
base one's arguments on average properties of the ensemble, such as
the typical number of emitters close to the edge. For example, instead of
suppressing radiation into $\Omega$ from gluons in the neighbourhood
of $\Omega$, one could just as well envisage a situation where the
neighbourhood of $\Omega$ is empty, automatically avoiding secondary
emissions into $\Omega$. We shall try to work through this argument with
the additional characteristic that we shall discretise the problem.
Taking $\Omega$ as a slice of width $\Delta \eta$ (sufficiently wide
that the two edges can be considered completely independent), the
probability of it staying empty down to some scale $t$ is $e^{-4\CF t
  \Delta \eta} \cS(t)$.  Let us suppose that the condition for secondary
radiation not to be emitted into $\Omega$ is determined by the
probability that bands of width $\delta \eta$ (`buffers') on either
side of the 
slice stayed empty down to a scale $t - \delta t$.  Then we have a
recurrence relation:
\begin{equation}
  \label{eq:BufferStep}
  e^{-4\CF t \Delta \eta} \cS(t) = e^{-4\CF [t \Delta \eta + 2(t-\delta
    t) \delta \eta] } \cS(t - \delta t)\,,
\end{equation}
where the term proportional to $\delta\eta$ in the exponent stems from the
suppression of primary emission in the `buffer' bands. Rewriting
\eqref{eq:BufferStep}  as a differential equation, we have
\begin{equation}
  \frac{\delta \ln \cS}{\delta t} = -8\CF t
  \left\langle \frac{\delta \eta}{\delta t} \right\rangle \,.
\end{equation}
The factor $\langle \delta \eta/\delta t \rangle $ can be
understood as the average `speed of motion' (probably proportional to $\CA$) of the
border between regions with and without emissions (\ie of the edge of
the buffer). The resulting form for $\cS$ is
\begin{equation}
  \label{eq:SWithBuffer}
  \cS(t) \sim e^{-4 \CF t^2 \left\langle \frac{\delta \eta}{\delta t}
    \right\rangle}\,,
\end{equation}
where we have (arbitrarily) assumed $\left\langle \frac{\delta
    \eta}{\delta t} \right\rangle$ to be independent of $t$.

\FIGURE{
    \epsfig{file=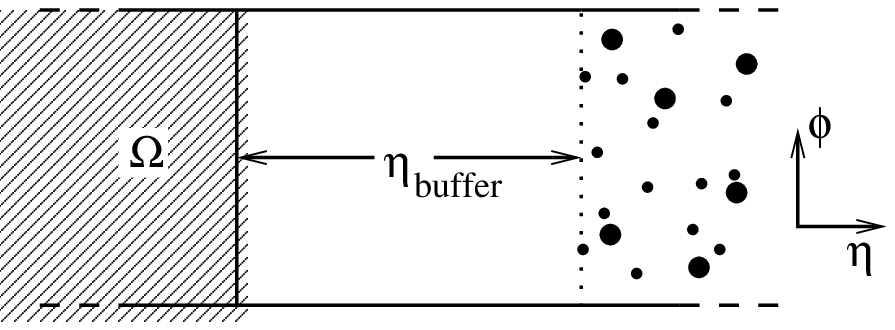,width=0.45\textwidth}
    \caption{A possible structure for emissions harder than some
      intermediate resolution scale $t'$ for an event in which
      $\Omega$ stays empty down to some scale $t > t'$. In this
      scenario, at scale $t'$ an empty buffer region exists between
      $\Omega$ and the emission closest to $\Omega$.}
    \label{fig:buffer}
}
One of the features of this mechanism is that resolving only
those emissions harder than some intermediate scale $t$', one will
see an extended empty buffer region surrounding $\Omega$, as in
figure~\ref{fig:buffer} (shown for one side of a wide slice). The
typical size of this buffer region should be of the order of
\begin{equation}
  \label{eq:buffer}
  \eta_\mathrm{buffer} \simeq (t-t') \left\langle \frac{\delta
      \eta}{\delta t} \right\rangle\,.
\end{equation}
It is the suppression of intermediate-scale primary radiation in the
buffer region which is responsible for the strong
suppression of $\cS(t)$ at large $t$, eq.~\eqref{eq:SWithBuffer}.

An important consequence of the existence of such a buffer region is that
the large-$t$ behaviour of $\cS(t)$ would be independent of the shape
and size of $\Omega$. This is because at large $t$ the edges of the
buffer region will be far from $\Omega$ and so the details of $\Omega$
can have no influence on the dynamics at the edge of the buffer. This
independence on the geometry of $\Omega$, together with the explicit
observation of a buffer region, would allow us to distinguish the
buffer mechanism from the alternative mechanism proposed at the beginning
of this subsection.

\subsection{Numerical results}

We mentioned above that there are two main problems in obtaining
all-order results. One is the complexity of the colour algebra in the
presence of large numbers of gluons. This can be eliminated by taking
the large-$\NC$ approximation, in which the squared matrix element
radiation can be broken down into a sum of independent terms each
associated with a different colour dipole \cite{BCM}.

The second source of complexity is the geometry of multi-gluon events,
which we treat using a Monte Carlo algorithm like that discussed
in \cite{dassalNG}, which essentially models a tree-like sequence
where a colour dipole emits a gluon, thus branching into two dipoles,
each of which may themselves go on to emit (there is a similarity to
the Ariadne event generator \cite{Ariadne}). However unlike a Monte
Carlo event generator such as Ariadne, our method has the property
that it gives results which are purely a function of $t$
(eq.~\eqref{eq:tdef}), and therefore contain just the piece which is
guaranteed to be correct, namely the leading logarithmic piece.

The Monte Carlo algorithm returns a function $\cS_\mathrm{MC}(t)$ in the
large-$\NC$ limit. We choose to correct this function so that at least
at order $\as^2$ the result is correct beyond the large-$\NC$
approximation. Accordingly, in what follows, we shall consider
\begin{equation}
  \cS(t) = [\cS_\mathrm{MC}(t)]^{\frac{2\CF}{\CA}}\,,
\end{equation}
rather than $\cS_\mathrm{MC}$ itself.

\FIGURE{
    \epsfig{file=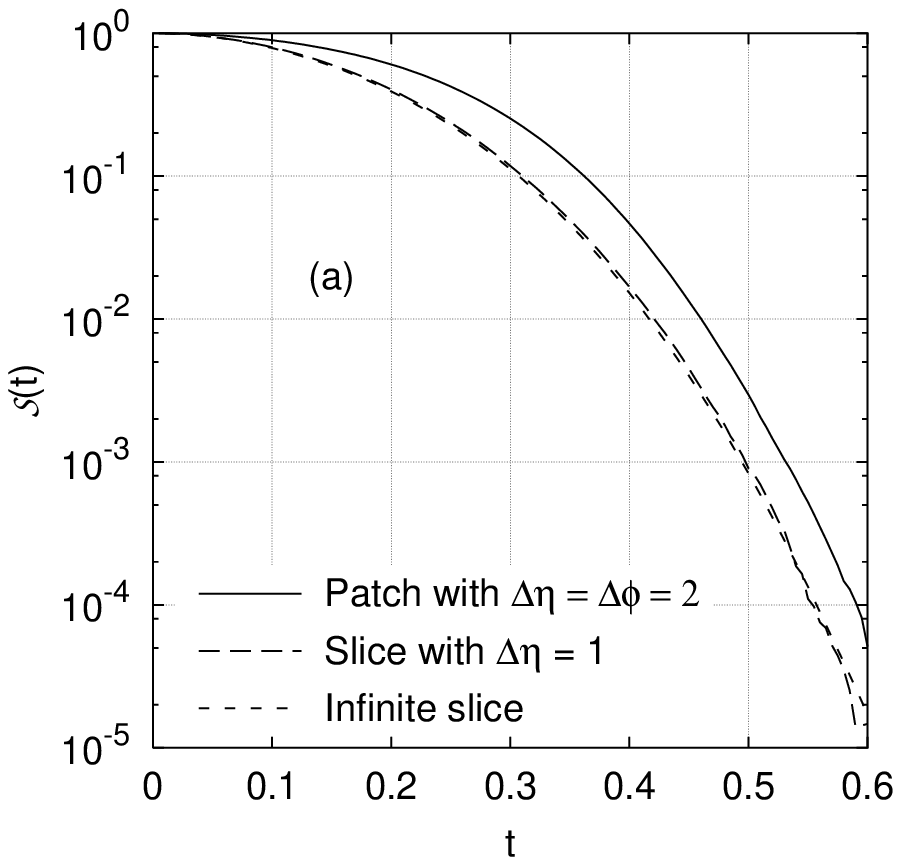,height=0.47\textwidth}\hfill
    \epsfig{file=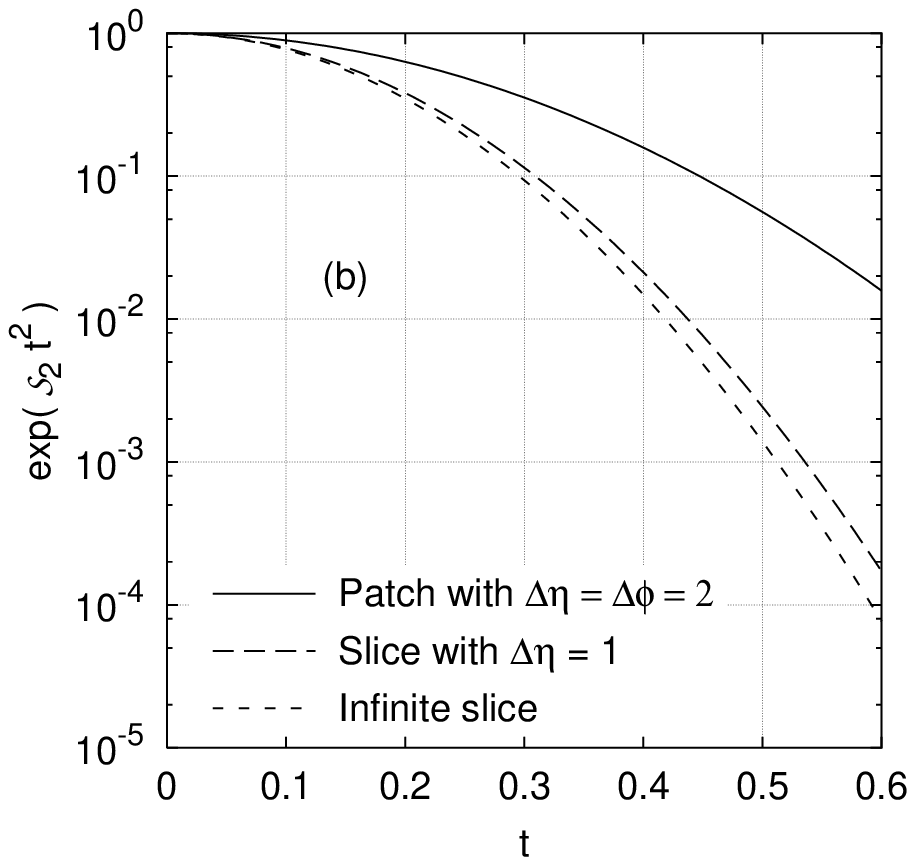,height=0.47\textwidth}
    \caption{A comparison between (a)  the full result for $\cS$ and (b) 
      the simple exponentiation of $\cS_2$. Beyond $t=0.5$ the
      differences between the curves for the finite and infinite
      slices are not significant compared to the statistical and
      systematic errors involved in the determination of $\cS$.}
    \label{fig:SAlone}
}

In figure~\ref{fig:SAlone}a we show the function $\cS(t)$ for three
different geometries for $\Omega$: a square patch in rapidity and
azimuth ($\Delta \eta=\Delta \phi=2$), a finite slice in rapidity
($\Delta \eta=1$) and an infinite 
slice in rapidity. We extend the $t$ scale beyond the
phenomenologically relevant region in order to better illustrate the
general features of $\cS$ in the different cases. The right-hand plot,
figure~\ref{fig:SAlone}b shows what would be obtained if there were a
simple exponentiation of the $\cS_2$ term, $\cS \simeq \exp(\cS_2 t^2)$.

There are various points to be noted: firstly, as $t$ increases, $\cS$
decreases with a behaviour roughly consistent with a Gaussian
suppression, at least for $t\lesssim0.5$. Secondly, at large $t$,
though the normalisation of
$\cS(t)$ depends on the geometry of $\Omega$, its behaviour in $t$
seems to be \emph{universal}.  This is to be compared to the geometry
dependence that would be present if the all-orders result stemmed from
a simple exponentiation of the $\cS_2$ term,
fig.~\ref{fig:SAlone}b. We note that up to $t\simeq0.5$, there is a
strong similarity between the actual $t$-dependence and that observed
in the exponentiation of $\cS_2$ for an infinite slice. Beyond
$t\simeq0.5$ however the suppression in the full calculation seems to
be grow much faster.

\FIGURE{
    \epsfig{file=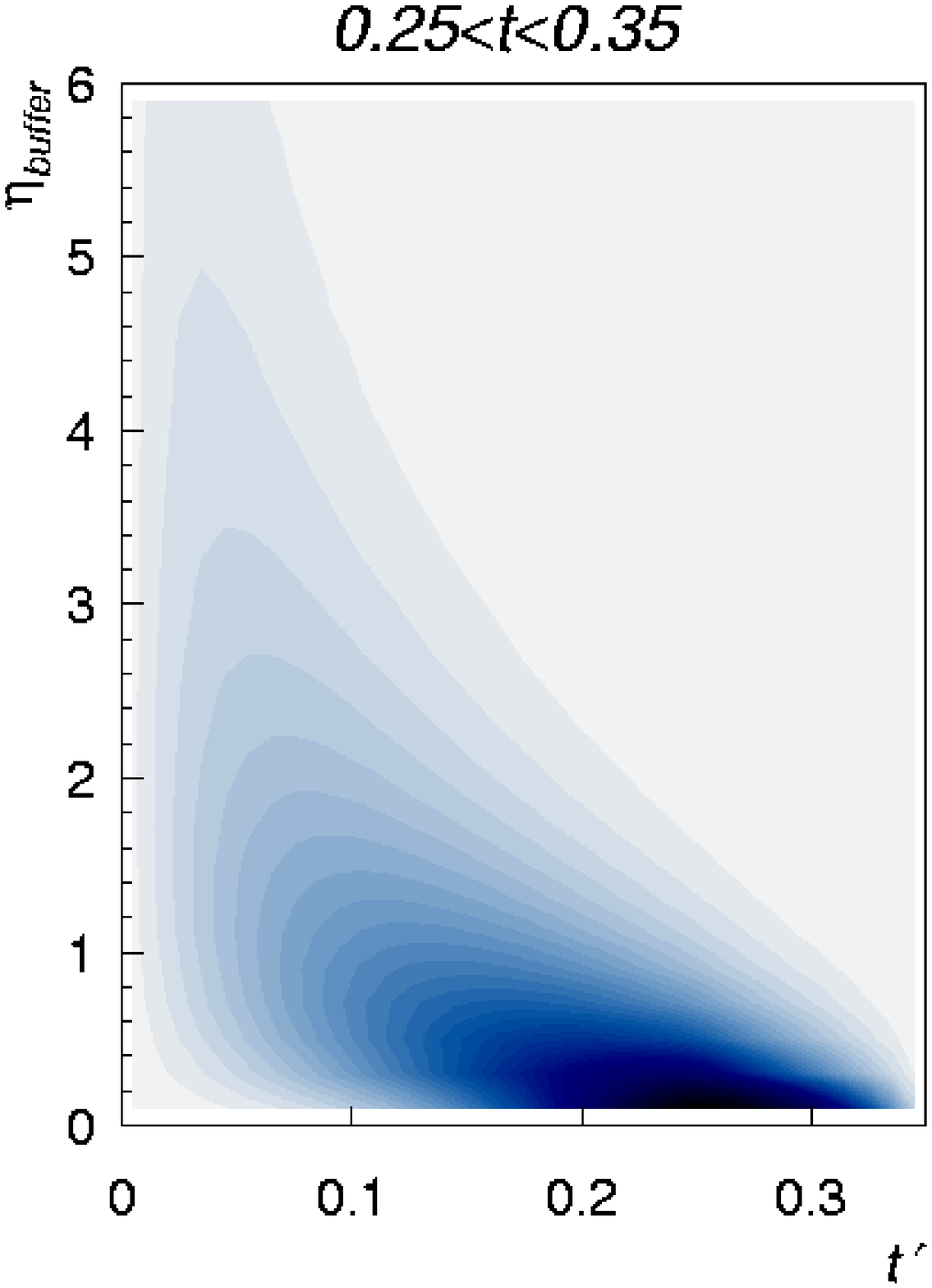,height=0.49\textwidth}\hfill
    \epsfig{file=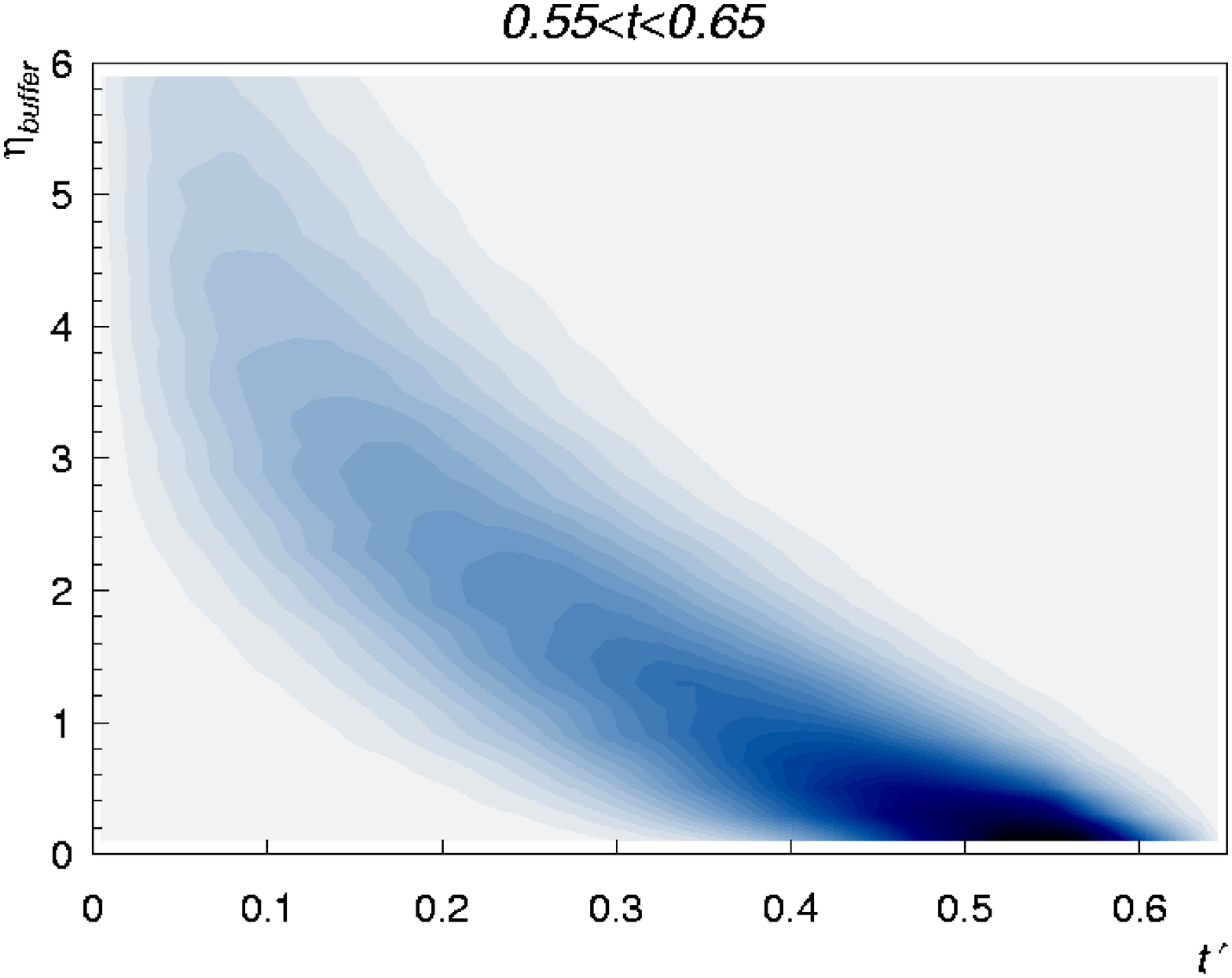,height=0.49\textwidth}
    \caption{Contour plots for the distribution of the size of the
      buffer region, $\eta_\mathrm{buffer}$, as a function of the
      intermediate resolution scale $t'$. For the larger $t$ range, the
      irregularities of the contours are an artifact due to limited
      statistics. The darkest contour corresponds to $\frac{1}{\sigma}
      \frac{d\sigma}{d\eta_\mathrm{buffer}}=2$.}
    \label{fig:contour}
}
The fact that the $t$-dependence of $\cS$ is the same regardless of the
geometry of $\Omega$ seems to suggest that the `buffer' mechanism
postulated above might well be responsible for the all-orders
behaviour of $\cS$. To examine this idea in more detail one can use the
Monte Carlo simulation to establish whether there really is a buffer region,
and to examine its size, $\eta_\mathrm{buffer}$, as a function of an
intermediate resolution scale $t' < t$. Figure~\ref{fig:contour} shows
contour plots for the distribution of $\eta_\mathrm{buffer}$ values as
a function of $t'$.  Here $\Omega$ is an infinite slice, and $t$ is
defined as the scale of the hardest secondary emission in $\Omega$
originating from coherent emission off partons which are to the right
of $\Omega$. Correspondingly the buffer region being considered is
that to the right of $\Omega$.

For the lower $t$ range the buffer region tends to be fairly small ---
for example at $t'=0.1$, the most likely buffer size is
$\eta_\mathrm{buffer} \simeq 0.7$.  However, increasing $t$ to $0.55 < t <
0.65$, the most likely size of the buffer region at $t'=0.1$ grows to
$\eta_\mathrm{buffer} \simeq 3.5$. Such an increase in buffer size for
fixed $t'$ as one increases $t$ is precisely what one would expect
from the `buffer' mechanism postulated earlier, eq.~\eqref{eq:buffer}.
From that equation 
one can also evaluate an effective value for $\langle \delta \eta
/\delta t \rangle \simeq 9$. Using eq.~\eqref{eq:SWithBuffer}, one
then finds that the coefficient of $t^2$ in the exponent is roughly
twice $\cS_2$ for an infinite slice.

This is not quantitatively consistent with what is seen in
figure~\ref{fig:SAlone}, however several points should be borne in
mind: (a) the arguments in section~\ref{sec:dynamics} are based on
average properties, whereas in the end we are interested in the
properties of rare events. So for example there is a certain spread in the
distribution of $\eta_\mathrm{buffer}$ and this can affect the
quantitative predictions by a non-trivial factor, as can the fact that
$\delta \eta/\delta t$ is itself only defined in an average sense
(b) one's estimate
for $\langle\delta \eta /\delta t\rangle$ depends somewhat on how
exactly one
deduces it from the plots, \eg whether as a derivative with respect to
$t'$ or $t$; this is connected to point (c), namely that our
assumption of a constant $\langle \delta \eta/\delta t\rangle$
(independent of $t'$)
may be an oversimplification. Indeed if $\langle \delta \eta/\delta
t\rangle$ is 
constant then eq.~\eqref{eq:buffer} one would expect the centre of the
distribution of $\eta_\mathrm{buffer}$ to be depend linearly on $t'$.
However in fig.~\ref{fig:contour} there seems to be some non-linear
dependence of the typical $\eta_\mathrm{buffer}$ on $t'$, though it is
not clear whether this is an artifact of $t$ not being sufficiently
asymptotic, or whether there is extra dynamics which remains to be
taken into account, such as a pile-up of emissions close to the
edge of the buffer. Furthermore at $t\simeq0.5$ the behaviour of
$\cS(t)$ departs from the initial approximate Gaussian, and starts to
fall much faster, also suggesting either that $t\lesssim 0.5$ is not
yet asymptotic, or that $\langle \delta \eta/\delta t\rangle$ has some
extra $t'$ 
dependence.

In conclusion, we believe that while we may have understood some of the
gross features of the dynamics involved in the all-orders behaviour of
$\cS(t)$, in particular the importance of a buffer mechanism in
determining the geometry-independent large-$t$ suppression of $\cS(t)$,
there are some significant details which remain to be
understood.

\subsection{Phenomenological implications}

\begin{figure}[htbp]
  \begin{center}
    \epsfig{file=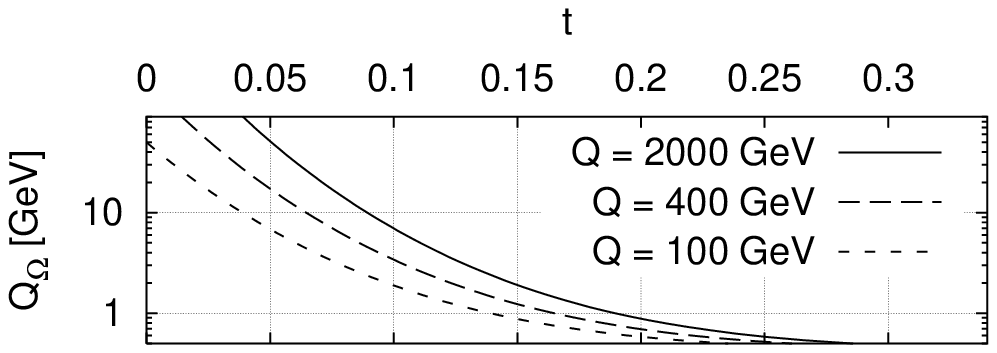,width=0.48\textwidth}\hfill
    \epsfig{file=t2q-many.eps,width=0.48\textwidth}\\
    \epsfig{file=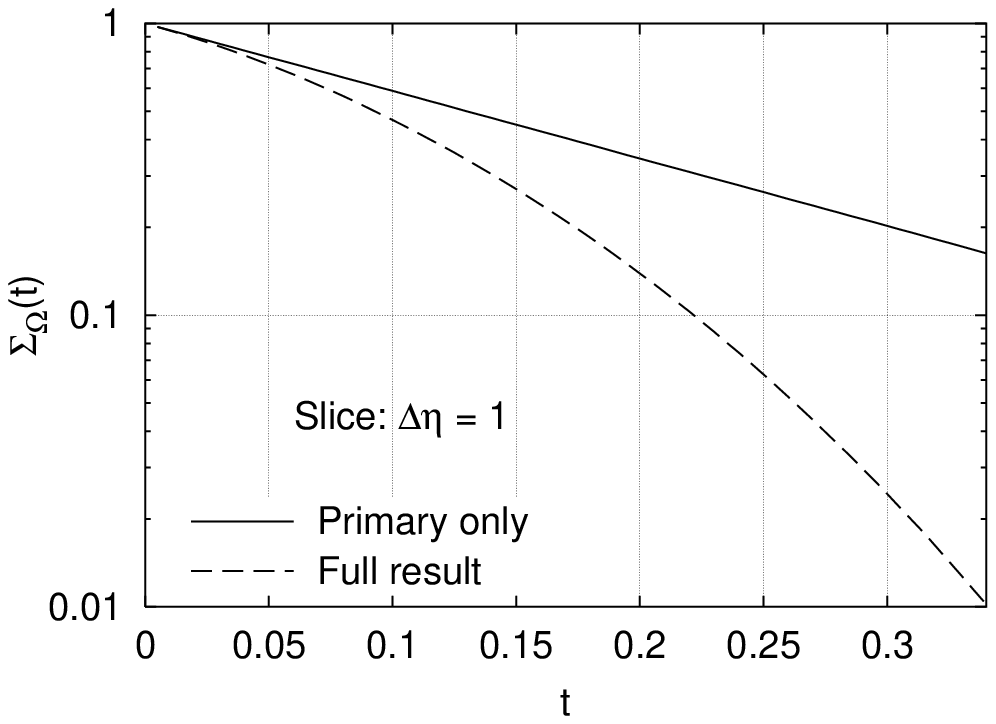,width=0.48\textwidth}\hfill
    \epsfig{file=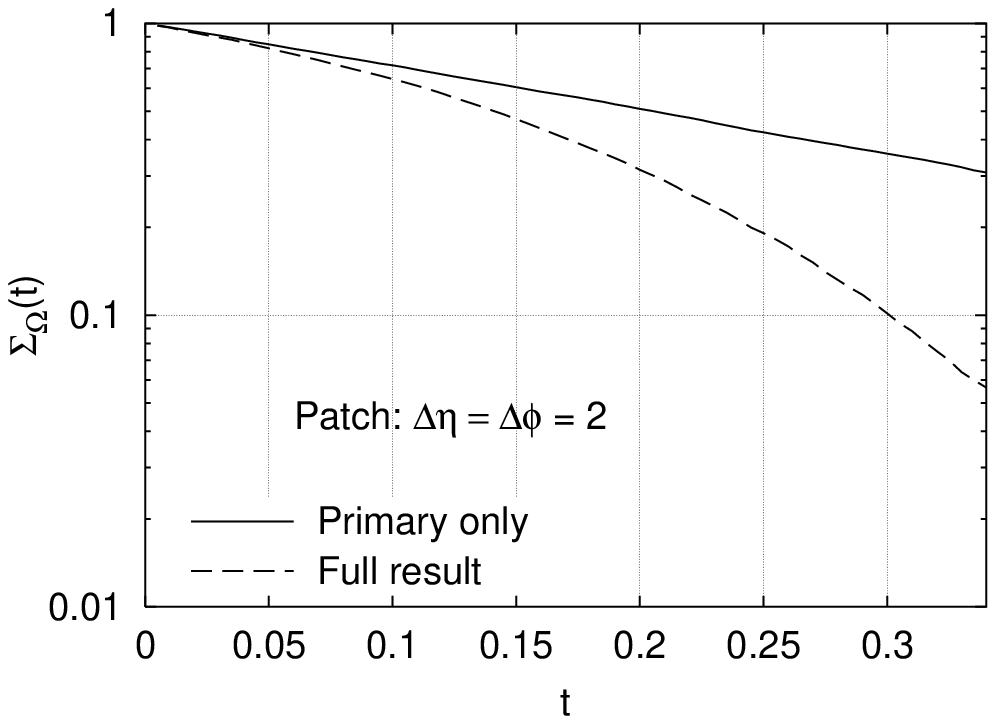,width=0.48\textwidth}
    \caption{$\Sigma_\Omega(t)$ for two different
      definitions of $\Omega$: in 
      the left-hand figure we consider the energy flowing into a slice
      of width $\Delta \eta = 1$, while in the right-hand figure we
      consider the energy flowing into a square patch in rapidity and
      azimuth, of size $\Delta \eta = \Delta \phi = 2$. The upper
      plots show the relation between $t$ and $Q_\Omega$ for different
      values of the centre-of-mass energy $Q$.}
    \label{fig:pheno}
  \end{center}
\end{figure}

To understand the phenomenological significance of the non-global
logarithms, it is interesting to compare results for $\Sigma(t(Q_\Omega,Q))$
with only primary emissions and with the full non-global
treatment. This is done 
in figure~\ref{fig:pheno} for two different geometries of $\Omega$, a
slice in rapidity and a square patch in rapidity and azimuth. 

The upper plots show how $t$ is related to $Q_\Omega$, for different values
of the centre of mass energy $Q$. Even at a very energetic
next-linear-collider, optimistically trusting the calculation down to
$Q_\Omega=0.5$~GeV one only just goes beyond $t=0.25$. For current energies
the largest value of $t$ that is accessible is in the range of $0.15$
to $0.2$. Taking $t=0.15$ as our reference value, the inclusion of the
non-global effects increases the suppression (on a logarithmic scale)
relative to that for just primary emissions by a factor of between
$1.5$ (patch) and $1.65$ (slice). To state it a different way,
ignoring non-global effects at $t=0.15$ overestimates the cross
section by between $30\%$ (patch) and $65\%$ (slice). At larger $t$
values, these figures rapidly become even more dramatic.

So the non-global effects are not only important from the point of
view of the formal correctness of the leading-log series, but also
numerically significant.

As an aside we comment on the feature of a divergence in the
distribution of $Q_\Omega$ at the Landau pole that has been observed in
certain circumstances in \cite{OdeSter,BKS}. Such a behaviour arises
when the suppression from $\Sigma_{\Omega,\cP}(t)$ is not sufficient
to compensate the divergence in $dt/dQ_\Omega$, \ie when
$\frac{\CF\cA_\Omega}{\pi \beta_0} < 1$. However the inclusion of
the non-global factor $S(t)$ ensures that $\Sigma_{\Omega}(t)$ always
goes to zero much faster than $dt/dQ_\Omega$ diverges.  Therefore
the distribution of $Q_\Omega$ should go to zero at the Landau pole
regardless of the size of $\frac{\CF\cA_\Omega}{\pi \beta_0}$.

Finally, we note that for a comparison to data it would also be
necessary to take into account non-perturbative effects. One way of
doing so might be in terms of power corrections
\cite{PowerCorrections}. For the mean $Q_\Omega$, in a normalisation in
which the power correction to the $e^{+}e^{-}$ thrust has a
coefficient $c_\tau = 2$, the coefficient for a region $\Omega$ would
be $c_\Omega = \cA_\Omega Q$ where $\cA_\Omega$ is the area of
$\Omega$. For the corresponding differential distribution one should
roughly expect a shift of the distribution by the same amount.

\section{Conclusions}

As has already been pointed out in \cite{dassalNG}, in the resummation
of any observable sensitive only to emissions in a limited angular region
$\Omega$ of phase space, there is a class of single-logs ---
`non-global logs' --- which leads to a breakdown of the picture of
independent primary emissions and strict angular ordering. That
picture has
been quite widely adopted in the literature, in certain instances
wrongly \cite{ADS,BG,BDMZ,OdeSter,BKS,Oderda}, at least to the accuracy that
was claimed.\footnote{There exist other cases,
  \eg \cite{CTTW,DLMSbrd}, whose final results are for global variables,
  but where intermediate steps of the derivation make reference to
  non-global hemisphere-variables without the inclusion of non-global
  logs. It is important not to use those results outside the context
  of the specific derivation for which they are intended without
  taking care of non-global effects as required.} %
To deal with this class of terms it is necessary to consider emission
from arbitrarily complicated ensembles of energy-ordered large-angle
gluons lying outside the region of sensitivity of one's observable.

In this article we have considered non-global logarithms for
observables such as the distribution of energy flow in restricted
angular regions between jets. This kind of measurement was originally
advocated, in \cite{MW,BKS}, for $2+2$-jet events in hadron-hadron
collisions. Here we have studied a simpler case, that of $2$-jet
events, in order to concentrate
on the specific features of these non-global logarithms, without the
difficulties that arise from the complex colour structure of four-jet
systems. We also point out that experimental studies of interjet
energy flows in $e^+e^-$ and DIS, insofar as they are free
from contamination by the underlying event, may provide useful
complementary information to that which could be extracted in
hadron-hadron colliders.

Our studies have led us to various conclusions. Most importantly
perhaps, is that non-global effects, as well as being formally of the
same order as those from primary emissions, are also numerically
almost as large.

More academically, one of the interesting aspects of non-global logs
that is specific to interjet energy flows, as opposed to the
single-hemisphere observables studied in \cite{dassalNG}, is that they
depend on the geometry of $\Omega$, the region in which one measures
the energy flow. At second order in $\as$ (the first order at which
these effects appear), for moderately sized patches, the magnitude of
non-global effects is roughly proportional to the perimeter of the
patch. In contrast, effects due to primary emissions scale as the area 
of the patch.

At all orders, we have only a large-$\NC$ numerical treatment for
non-global effects.
Despite this it is possible to obtain some interesting insights into
the mechanisms that 
are of relevance. One notable result is that, modulo a
geometry-dependent normalisation, the asymptotic $t$-dependence of the
non-global suppression factor $\cS$ seems to be independent of the
size and shape of $\Omega$. This fact, together with
figure~\ref{fig:contour}, lends support to the hypothesis that at
intermediate scales there is an empty `buffer region' around $\Omega$
and that a significant part of the non-global suppression factor $\cS$
stems from the suppression of intermediate-scale primary radiation in
the buffer region. This allows us to postulate that in other
processes, such as $2+2$-jet production at
hadron colliders, one will see similar results, with the asymptotic
$t$-dependence of $\cS$ being given by that of
figure~\ref{fig:SAlone}a, raised to a power which depends on the number
of jets and whether they are quark or gluon jets.

\acknowledgments

We wish to thank Yuri Dokshitzer, George Sterman and Bryan Webber for
helpful discussions. One of us (MD) gratefully acknowledges the hospitality of 
the LPTHE where part of this work was carried out.


\end{document}